\documentclass[aps,pra,10pt,twocolumn,superscriptaddress,floatfix]{revtex4-1}

\usepackage[utf8]{inputenc}
\usepackage[T1]{fontenc}
\usepackage{bm}
\usepackage[english]{babel}	
\usepackage{breakcites}
\usepackage{lmodern}
\usepackage[normalem]{ulem}

\usepackage{ae,aecompl}
\usepackage{subfigure}
\usepackage{latexsym,stmaryrd}
\usepackage{graphicx}
\usepackage[labelfont=bf,font={small},labelsep=period]{caption}
\usepackage[hidelinks,linkcolor=blue,colorlinks=true,urlcolor=blue,citecolor=blue]{hyperref}
\usepackage{braket}
\usepackage{amsmath}
\usepackage{upgreek}
\usepackage{cleveref}
\usepackage{comment}
\usepackage{url}
\usepackage{amssymb}

\newcommand{\micro}{$\mathrm{\mu}$}

\newcommand{\pin}{P_\mathrm{inj}}
\newcommand{\lin}{\lambda_\mathrm{inj}}

\begin{document}

\title{Probing injection locking of an integrated nanolaser pair through RF beating}

\author{Jan-Niklas Topf}
\affiliation{Université Côte d’Azur, CNRS, Institut de physique de Nice, Nice, France} 

\author{Roberto Gajardo}
\affiliation{Université Côte d’Azur, CNRS, Institut de physique de Nice, Nice, France}

\author{Guillaume Huyet}
\affiliation{Université Côte d’Azur, CNRS, Institut de physique de Nice, Nice, France}

\author{Fabrice Raineri}
\affiliation{Université Côte d’Azur, CNRS, Institut de physique de Nice, Nice, France}
\affiliation{Institut Universitaire de France, Paris, 75231, France}

\author{Guilhem Madiot}
\affiliation{Corresponding author: guilhem.madiot@univ-cotedazur.fr}
\affiliation{Université Côte d’Azur, CNRS, Institut de physique de Nice, Nice, France}

\begin{abstract}
We demonstrate simultaneous optical injection locking of an integrated pair of photonic-crystal nanolasers coupled to a common silicon waveguide. Due to fabrication-induced disorder, the two nanolasers exhibit distinct emission wavelengths, allowing their individual locking ranges to be resolved as a function of the injection-laser detuning and power. We identify a region where both nanolasers are simultaneously locked to the master laser. To probe this collective locked state, the two nanolasers are independently pump-modulated at distinct radio frequencies. A beat note at the difference frequency appears only within the overlap region, providing a direct RF signature of simultaneous locking. These results establish RF beating as a compact diagnostic of common-master injection locking in integrated nanolaser systems.
\end{abstract}

\maketitle

\section{Introduction}










Networks of coupled laser oscillators are being explored as physical substrates for computation and signal processing. They can implement binary phase states relevant to Ising-type spin models \cite{Utsunomiya2015BinaryPhase}, emulate coupled phase oscillators such as the Kuramoto model \cite{Takemura2021Kuramoto}, and form coherent Ising machines out of mutually coupled fiber lasers \cite{babaeian_single_2019}. 
More generally, controlling a laser's phase, frequency, or dynamical state through injected light enables applications ranging from neuromorphic photonics \cite{puts_all-optical_2025,pappas2024ultra} and bandwidth-enhanced optical sources \cite{Chen2011PhCInjection} to coherent beam combining and phased-array architectures \cite{ nguyen_improving_2022}.  Across these applications, a well-defined phase relationship must be imposed and maintained between laser oscillators that would otherwise run independently.

Optical injection locking offers a direct route to such synchronization. Seeding a slave laser with light from a master source pulls the slave's frequency and phase toward the injected field over a finite range of detuning and injection strength, defining a locking region known as an Arnold tongue \cite{Liu2020InjectionLocking,Lau2009InjectionTutorial}. Historically used to reshape the intensity- and phase-modulation response \cite{Lidoyne1991Modulation,Kikuchi1984Amplitude} and to narrow the linewidth \cite{Liu2020InjectionLocking} of semiconductor lasers, 
related injection-locking schemes have also been explored in mutually coupled, monolithically integrated laser arrays, where mutual injection can enhance the modulation response \cite{Sun2015MutualInjection} and where the coupled dynamics depend strongly on the phase delay between cavities \cite{Noda1990MutualInjection}.
On photonic-crystal platforms, injection locking of a single nanolaser has been used to raise the modulation bandwidth of directly modulated sources for optical interconnects \cite{Chen2011PhCInjection} and to induce bistable switching \cite{Bazin2012Bistable} and flip-flop memory operation \cite{Fitsios2016PhCMemory} in hybrid III-V on silicon nanolasers. 
More recently, injection locking has also been used to synchronize the self-pulsing dynamics of two nanocavities coupled on a photonic integrated circuit \cite{beltramo2025injection}.

Photonic-crystal nanolasers integrated on silicon waveguides combine a wavelength-scale footprint with 
compact integration, making them attractive building blocks for dense, chip-scale laser networks. 
Coupled photonic-crystal nanolasers already exhibit a variety of collective behaviors, including spontaneous mirror-symmetry breaking \cite{Hamel2015SymmetryBreaking}, mesoscopic limit cycles \cite{Marconi2020LimitCycles}, and directional emission engineered around exceptional points \cite{madiot2024harnessing}. In each of these cases, however, the collective state arises from the mutual coupling between the lasers themselves, rather than from an external field imposing a common phase and frequency on the whole array. 
An external master field instead provides a common phase and frequency reference while preserving the possibility of addressing the individual nanolasers. Simultaneous optical injection locking of distinct, frequency-disordered photonic-crystal nanolasers to a common waveguide-injected master remains unexplored experimentally.

\begin{figure*}
    \centering
    \includegraphics{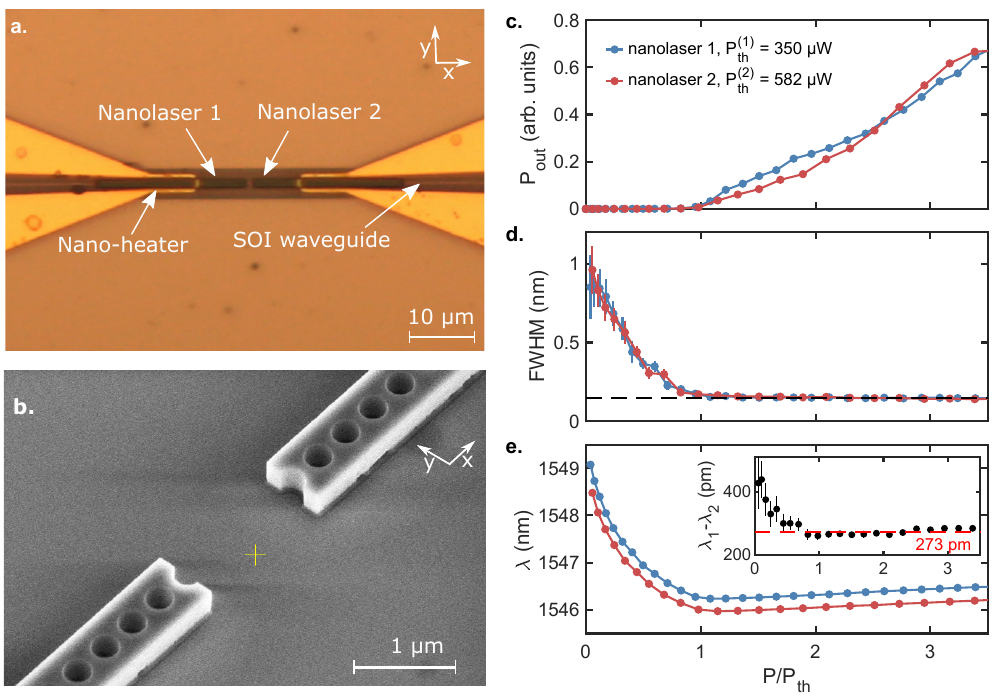}
    \caption{
    \textbf{a.} Optical microscope image of the system, highlighting two III-V nanolasers on top of an SOI waveguide, and with gold thermo-optic nano-heaters \textbf{b.} Scanning-electron micrograph of the nanolasers' terminations. The image is taken prior to the encapsulation step. The waveguide is hidden by the intermediate silica layer \textbf{c.} Integrated spectral power of the waveguide output , \textbf{d.} emission linewidth (FWHM), and \textbf{e.} emission wavelength $\lambda$, as a function of the pump power normalized to threshold, $P/P_\mathrm{th}$ with $P^{(1)}_\mathrm{th} = 350~\text{\textmu W}$ and $P^{(2)}_\mathrm{th} = 582~\text{\textmu W}$.  The dashed line in (d.) indicates the spectrometer resolution (150 pm). Inset : wavelength difference $\lambda_1 - \lambda_2$, converging to approximately $270~\mathrm{pm}$ above threshold. 
The error bars indicate the 95\% confidence interval.}
    \label{fig1}
\end{figure*}

Here, we injection-lock two waveguide-coupled photonic-crystal nanolasers to a common external master laser, injected directly into the bus waveguide that couples them. By sweeping the injection wavelength and power, we map the individual locking range of each nanolaser and identify the region in which both ranges overlap and the two lasers lock simultaneously to the same master field. Modulating the pump of each slave laser at a distinct frequency, we find that a beat note appears at the difference frequency only when both nanolasers are locked, giving a direct radio-frequency signature of their coherent, common-mode operation. A coupled-mode description of the injected, waveguide-coupled array accounts for this behavior and outlines that the beat-note amplitude depends jointly on the photon numbers and the relative phases of the two cavity fields, rather than on their intensities alone.

\section{Results}
We study two nanolasers integrated on a silicon-on-insulator (SOI) waveguide. The nanolasers are based on one-dimensional photonic crystal cavities approximately 50~\micro m long, 250~nm wide, and 300~nm high. They are fabricated from an indium phosphide membrane containing four InGaAsP quantum wells that serve as the active gain medium. The coupling between each nanolaser cavity and the underlying silicon waveguide is determined by both the waveguide geometry and the cavity-to-waveguide separation. The latter is fixed by a $250~\text{nm}$-thick SiO$_2$ spacer deposited by PECVD between the InP layer and the silicon waveguide. For the devices investigated here, the waveguide width is $400~\text{nm}$, corresponding to a coupling quality factor of approximately $Q_c \approx 4000$. Although the cavities are designed with identical nominal geometries, fabrication imperfections lead to parametric disorder in the nanolaser array. The laser cavities are pumped using a continuous-wave laser with a wavelength of $\lambda_{\text{pump}} = 1064~\text{nm}$ at room temperature. 

We compare the nanolasers' individual responses in \cref{fig1}. We increase the optical pump power on a given nanolaser while the other is off, and measure the output spectrum with a spectrometer coupled to a cooled near-infrared camera. We fit the spectra with a Lorentzian lineshape and plot the integrated power (a.), the emission linewidth (b.), and the emission wavelength (c.). All quantities are plotted as a function of each device's pump power, normalized to the power at threshold, $P_{\text{th}}$. We extract lasing thresholds of $P_{\text{th}} = 350~\text{\textmu W}$ and $582~\text{\textmu W} $ for nanolasers 1 and 2, respectively. Above threshold the two emission wavelengths stabilize with a residual detuning of $\lambda_1 - \lambda_2 \approx 270~\mathrm{pm}$ (inset of \cref{fig1}c), which sets the natural frequency mismatch that the injected master field must overcome to synchronize the array. We mainly attribute the difference between the threshold powers to asymmetric optical alignment of the pump beams, resulting in different absorption rates of the pumps by the nanolasers. 

Once normalized, we observe two similar input-output curves, with only small differences in the output power, and extremely similar optical linewidths, narrowing from approximately $1~\text{nm}$ down to the spectrometer resolution at threshold, at 150 pm. The emission wavelengths follow the same evolution, i.e. a blueshift below threshold -- associated with free-carrier effects -- followed by a redshift above threshold that arises from a thermo-optic effect as the photon population increases while the carrier density remains clamped.

\begin{figure*}[!ht]
    \centering
    \includegraphics{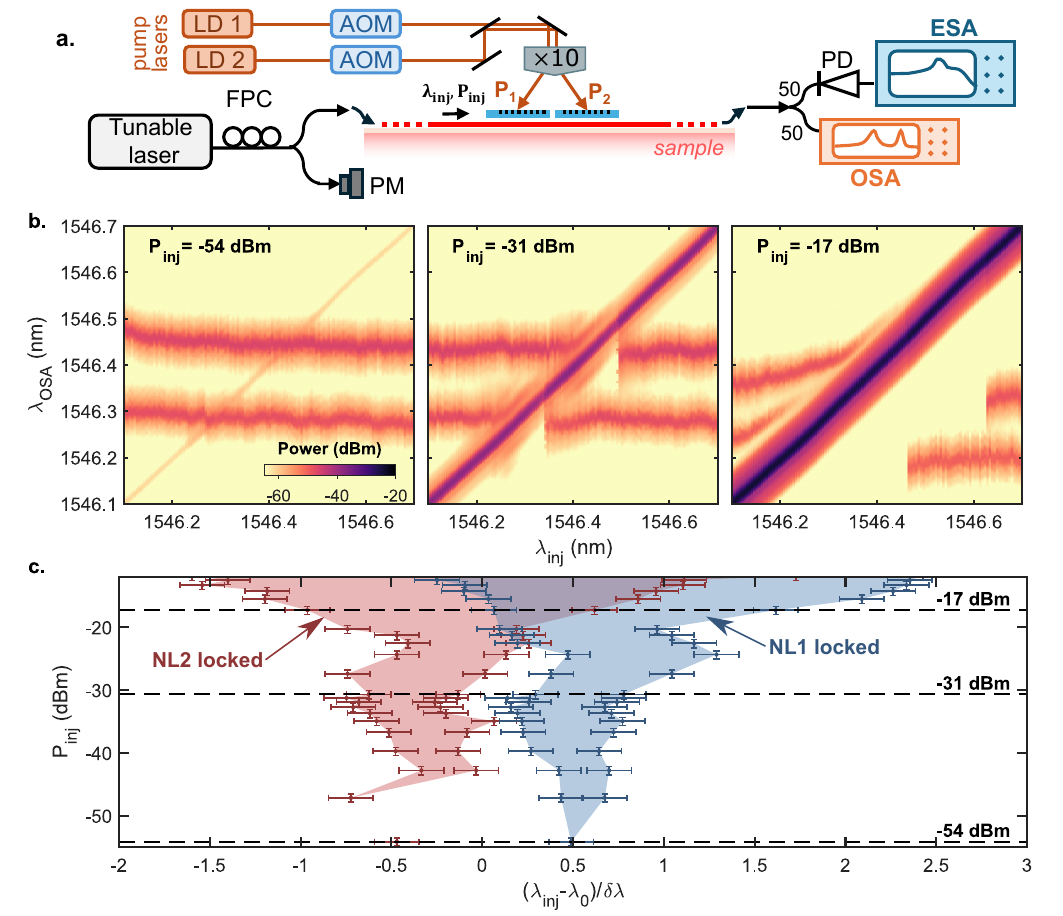}
    \caption{\textbf{Injection locking of the nanolaser array.}
    \textbf{a.} Schematic of the experimental setup. 
    \textbf{b.} OSA spectral maps recorded while sweeping the injection wavelength
    $\lin $, for three injection powers $\pin  = -54$,
    $-31$, and $-17~\text{dBm}$ (left to right). The color encodes the detected
    optical power (dBm); the bright diagonal is the injected master line and the
    horizontal features are the two free-running slave lasers.
    \textbf{c.} From the above maps, the spectral positions at which each nanolaser locks or unlocks to the injection are reported in the parameter space $\{\lin,\pin\}$, defining Arnold's tongues for each nanolaser (NL1 in red, NL2 in blue).
    Global locking is observed when the tongues overlap, above $\pin  \approx -20~\mathrm{dBm}$.
    Dashed lines indicate the three powers shown in (b.). Error bars along x and y correspond to the OSA resolution and the power calibration uncertainty, respectively.}
    \label{fig2}
\end{figure*}

Both nanolasers, coupled to the same waveguide, are now pumped above threshold simultaneously by two independent pump lasers; the corresponding schematic setup is shown in \cref{fig2}~a). In the master-slave configuration, these two nanolasers act as the slave lasers. The pump powers are controlled by two acousto-optic modulators (AOMs) and directed onto the individual nanolasers through free-space optics and a $\times10$ microscope objective. A tunable laser, coupled into the waveguide via fiber-to-chip coupling, serves as the master laser. Between the master laser and the sample, a polarization controller and an optical isolator ensure efficient coupling to the chip and suppress back-reflections that could otherwise perturb the master laser's operation. A beam splitter is placed in the same path, with one output sent to a power meter and the other to the sample, allowing the injected master power to be calibrated. The system output is likewise extracted by fiber-to-chip coupling and sent to a 50/50 splitter: one output is connected directly to an optical spectrum analyzer (OSA), while the other is routed to a photodiode (PD) feeding an electrical spectrum analyzer (ESA). 

To perform injection locking on the nanolaser array, the wavelength of the master laser $\lin $ is swept across the emission wavelengths of the slave lasers, from $1546.1$ to $1546.7~\mathrm{nm}$, at several injection powers $\pin $, while the output spectrum of the device under test (DUT) is recorded with the OSA. Representative measurements are shown in \cref{fig2}~b), where the master-laser power is set to $-54$, $-31$, and $-17~\mathrm{dBm}$ from left to right, respectively. Each map plots the OSA spectral power as a function of the injected master wavelength. At the lowest injection power of $-54~\mathrm{dBm}$, the master laser sweeps across the two slave lasers with essentially no locking: the slave-laser wavelengths remain unchanged. At $-31~\mathrm{dBm}$, the slave lasers lock to the master laser one at a time and then jump back to their original emission wavelengths; here the locking range is only wide enough to capture one laser at a time. At $-17~\mathrm{dBm}$, the master laser successively locks the first slave laser, then both slave lasers simultaneously, and finally only the second slave laser. This behavior is summarized in \cref{fig2}~c), where $\pin $ is plotted against the detuning of the master laser from the free-running nanolaser, i.e. how far the injected wavelength lies from the average slave lasers' natural wavelength $\lambda_0$, normalized by the free-running detuning $\delta\lambda = \lambda_1 - \lambda_2 \approx 270~\mathrm{pm}$. The data form two locking regions -- or Arnold's tongues -- whose width increases with $\pin $; the tongues begin to overlap at an injection power of $\pin  \approx -20~\mathrm{dBm}$, and within this overlap region both slave lasers are locked by the same master laser.

\begin{figure}
    \centering
    \includegraphics{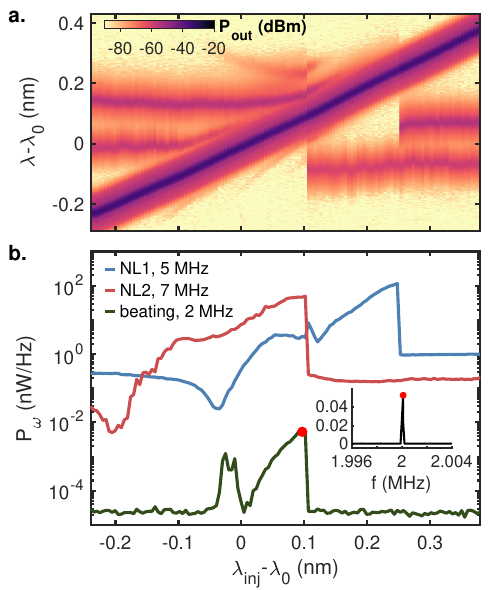}
    \caption{\textbf{Simultaneous locking probed by pump modulation.}
Data recorded at a fixed injection power of, with the slave-laser pumps amplitude-modulated at $5~\mathrm{MHz}$ (nanolaser~1) and $7~\mathrm{MHz}$ (nanolaser~2).
\textbf{a.} OSA spectral map versus master wavelength $\lin $; the color encodes the detected optical power (dBm). 
\textbf{b.} Power spectral density of the output signal at 5 MHz (blue), at 7 MHz (red), and at the beat-note (2 MHz, black), 
as a function of $\lin $. A $2~\mathrm{MHz}$ beat note appears only where both slave lasers are simultaneously locked to the master, providing an RF signature of coherence between the nanolasers. Inset: a typical RF spectrum centered at the beating, $f=2$ MHz. The peak amplitude is used to construct the plots in \textbf{b.}.}
    \label{fig3}
\end{figure}

To gain more insight into the regime where both slave lasers are locked, the AOMs are used to apply amplitude modulation to the pump of each slave laser. Nanolaser~1 is modulated at $5~\mathrm{MHz}$ and nanolaser~2 at $7~\mathrm{MHz}$.  The same master-laser wavelength sweep is then performed, and an ESA spectrum is recorded at every wavelength step. \cref{fig3}~b) plots the photocurrent power spectral density $P_\omega$ measured at three fixed analysis frequencies: $5~\mathrm{MHz}$ (blue) and $7~\mathrm{MHz}$ (red), corresponding to the modulation frequencies of the two slave lasers, and $2~\mathrm{MHz}$ (black), the difference frequency at which the beat note between the two modulated slave lasers is expected. In the inset of \cref{fig3}.b, we show a typical RF spectrum centered at the beat note frequency, obtained for $\lambda_\text{inj} - \lambda_0 = 97$ pm.

Within its locked regime, each slave laser produces an RF tone $P_\omega$ at its modulation frequency that increases as the master-laser wavelength is swept. In the region where the two locking ranges overlap, the beat note at $2~\mathrm{MHz}$ becomes visible, whereas it is absent when none or only one of the slave lasers is locked to the master laser.

The dynamics of each injection-locked nanolaser $j$ is described by the field rate equation \cite{Seegert2026}
\begin{equation}
\frac{dE_j}{dt}
=
\left[-\Gamma_j+G_j(1+i\alpha)-i\Delta_j\right]E_j
+\kappa_j E_{\mathrm{in}},
\label{eq:injected_laser}
\end{equation}
coupled to the carrier dynamics, with the optical pump weakly modulated at frequency $\omega_j$. Here $E_j$ is written in the reference frame rotating at the injection frequency $\omega_{\mathrm{in}}$, $\Delta_j$ is the free-running detuning from the injection, $\Gamma_j$ is the cavity loss rate, and $G_j$ is the carrier-dependent gain. The coefficients $\kappa_1=i\sqrt{\Gamma_c}$ and $\kappa_2=i\sqrt{\Gamma_c}e^{-i\phi}$ describe the coupling of nanolasers 1 and 2 to the output waveguide, respectively, where $\Gamma_c=\omega_0/(2Q_c)$ is the cavity amplitude decay rate into the waveguide and $\phi$ is the propagation phase accumulated in the waveguide between the two cavities \cite{madiot2024harnessing}. In the injection-locked regime, Eq.~(\ref{eq:injected_laser}) admits a stationary solution $E_{j,0}$ with a constant phase relative to the injected field. A small modulation of the pump then produces a field perturbation
\begin{equation}
E_j(t)=E_{j,0}
+\delta E_j(\omega_j)e^{i\omega_j t}
+\delta E_j(-\omega_j)e^{-i\omega_j t},
\end{equation}
whose amplitude is determined by the small-signal response of the locked state. 

When both nanolasers are locked, their optical frequencies are identical.
The optical field collected at the waveguide output results from the coherent
superposition of the injected field and the fields emitted by the two
nanolasers,
\begin{equation}
E_{\mathrm{out}}(t) = E_{\mathrm{in}} + i\sqrt{\Gamma_c}\Big( E_1(t) + e^{-i\phi}E_2(t)\Big),
\end{equation}
In the injection-locked regime, the field of each nanolaser is
stationary in the reference frame of the injected field. 
The photodiode measures the square modulus of the output field, $P_{\mathrm{out}}(t)\propto |E_{\mathrm{out}}(t)|^2$.
The beating of each modulated nanolaser field with the stationary optical fields therefore produces RF components at $\omega_1$ and $\omega_2$.
Moreover, the cross product between the modulated components of the two nanolasers generates a component at $\Omega = |\omega_1-\omega_2|$ :
\begin{equation}
\Gamma_c e^{i\phi}
\delta E_1(\omega_1)\delta E_2^*(\omega_2)
e^{i\Omega t}+\mathrm{c.c.}
\end{equation}

This term appears at the difference frequency only when both nanolasers are locked to the injected field. Indeed, if nanolaser $j$ is unlocked, its field is no longer stationary in the injection reference frame and acquires a residual phase drift. The corresponding cross term is then shifted.

Importantly, $P_{\omega}$ depends on both the magnitude and the phase of the complex cross-products entering its expression. The modulation response is therefore not determined solely by variations of the intracavity photon numbers. Consequently, $P_{\omega}$ is intrinsically sensitive to both the photon numbers and the phases of the cavity fields, making the RF response a direct signature of the coupled amplitude--phase dynamics of the system.

\section{conclusion}

In conclusion, we have demonstrated the simultaneous optical injection locking of two frequency-detuned photonic-crystal nanolasers to a common master field injected through a shared waveguide. By mapping their individual Arnold tongues, we identify the parameter region in which both nanolasers are simultaneously locked to the same optical reference. Independent modulation of the two optical pumps reveals a beat note at the difference frequency only in this overlap region, providing an RF signature of the common locked state. The corresponding coupled-mode description shows that this RF response depends on both the amplitudes and the relative phases of the nanolaser fields. These results establish common-master injection locking as a simple route toward the coherent control and synchronization of integrated nanolaser networks.

In a further study, synchronous detection of both the amplitude and phase of each RF signal could provide access to the intracavity field intensities (photon numbers) and associated 
phases. Applied to larger nanolaser arrays under external optical injection, this approach could help distinguish partial from global locking and provide a useful diagnostic of complex spatiotemporal dynamics.

\begin{acknowledgements}
The authors acknowledge I.Sagnes, G.Beaudoin, and K.Pantzas for growing the InP based materials used in this work. 

This work was partly supported by France 2030 government grants (ANR-22-PEEL-0010), by the French National Research Agency (ANR), Grants No ANR-25-CE24-1834 and ANR-23-CE24-0013, and by the RENATECH network. 
\end{acknowledgements}

\bibliography{refs}

\end{document}